# Knowledge-aware Autoencoders for Explainable Recommender Sytems

Vito Bellini*, Angelo Schiavone*, Tommaso Di Noia*,
Azzurra Ragone•, Eugenio Di Sciascio*
*Polytechnic University of Bari
Bari - Italy
firstname.lastname@poliba.it
•Independent Researcher
azzurra.ragone@gmail.com

## ABSTRACT

Recommender Systems have been widely used to help users in finding what they are looking for thus tackling the information overload problem. After several years of research and industrial findings looking after better algorithms to improve accuracy and diversity metrics, explanation services for recommendation are gaining momentum as a tool to provide a human-understandable feedback to results computed, in most of the cases, by black-box machine learning techniques. As a matter of fact, explanations may guarantee users satisfaction, trust, and loyalty in a system. In this paper, we evaluate how different information encoded in a Knowledge Graph are perceived by users when they are adopted to show them an explanation. More precisely, we compare how the use of categorical information, factual one or a mixture of them both in building explanations, affect explanatory criteria for a recommender system. Experimental results are validated through an A/B testing platform which uses a recommendation engine based on a Semantics-Aware Autoencoder to build users profiles which are in turn exploited to compute recommendation lists and to provide an explanation.

## 1 INTRODUCTION

In recent time we assisted to the rising of Deep Learning models in many fields such as Computer Vision, Speech Recognition, Natural Language Processing and, very recently, few attempts have also been made to solve the Recommendation problem. Deep Learning techniques have proven their strength thus gaining the attention of both researchers and companies and being widely deployed in nowadays recommender systems. While research has mainly focused on improving accuracy metrics in recommenders, under the hood, their algorithms are becoming more and more complex thus making extremely hard to understand the reasons behind model predictions for a particular input. This recently led both researchers and companies to pay more attention to explainable models. Indeed, it has been proven that showing to users an explanation for the provided recommendation leads to better interaction with the recommender system. Moreover, when users understand how the system works, they can refine their preferences in order to get a better recommendation according to their tastes. However, in many popular recommenders such as Amazon or Netflix, the explanation provided is still very poor, as it is essentially based on popularity basis: it just tells that users with similar tastes have enjoyed the suggested items. It turns out that this kind of explanation is not perceived as a valid justification of why the system is recommending certain items and it hardly improves users loyalty in the system. On the other hand, a content-based explanation turns out to be more engaging from the user's point of view because it makes users aware about item's attributes that might be relevant for them.

Furnishing a content-based explanation seems to be much more difficult because item descriptions are not always available and they are not easy to maintain. Thus, some attempts have been made in order to exploit Knowledge Graphs as data source for items' content description. Generally, KGs such as DBpedia[1] or Wikidata[2], provides huge information of different types and then finding what works better to build an explanation is a difficult task for computer agents.

In this paper we propose to exploit a Semantics-Aware Autoencoders (SemAuto) [2] to compute explainable recommendations. Originally developed to cope with the cold start problem, in SemAuto the structure of the DBpedia KG is injected within an Autoencoder Neural Network, whose structure is built by mimicking the existing connections in the KG. Then, after feeding such a network with user ratings, weights associated to the hidden neurons are extracted and eventually used to build knowledge-aware user profiles which are eventually used to compute recommendations. In

[1] http://dbpedia.org
[2] https://wikidata.org

[3] we prove that SemAuto can also be effectively used to compute recommendations in non-cold situations reaching very competitive results in terms of accuracy and diversity.

Here we show how the model built by SemAuto can also be adopted to compute content-based explanations to recommended items. We evaluated the effectiveness of our approach through an A/B testing platform with 892 volunteers and compared its results to two baselines. We tested both a pointwise and a pairwise explanation style by exploiting different kinds of information on DBpedia (categorical and factual), in order to investigate how the effectiveness of the proposed explanation changes according to the selected properties. The main research questions we address in this paper are then:

**RQ1** Can we assume that the information encoded in the hidden layer of the SemAuto autoencoder is representative of user preferences?

**RQ2** Given a content-based explanation built upon the SemAuto model, is a pairwise explanation better than a simple point-wise one for the user?

The remainder of the paper is structured as follows: in the next section we recap the most prominent works related to the explanation in the recommendation scenario. In Sections 3 we provide an overview of Knowledge Graphs and Semantics-Aware Autoencoders. We describe how our experiments were conducted in Section 4 and discuss the results in Section 5. Conclusion and future work close the paper.

## 2 RELATED WORKS

Making a Recommender System (RS) transparent to users is getting more and more relevance since it may lead to users retain [8]. Different studies [15, 18] have pointed out that introducing transparency in the recommendation process may have lots of advantages because users appear to be more satisfied with the recommendation if they are aware of the reasons why certain items are suggested. Furthermore, the provided explanation may also convince users to try items they would have normally ignored, thus improving users confidence in the system.

Since the explanation may be decoupled from the recommendation process, a distinction between *transparency* and *justification* has to be made [20]. The explanation brings *transparency* to the system if it makes users aware about how the recommender engine works, explaining somehow the underlying algorithm behind the proposed suggestions. This is usually the case of those explanations computed along with the recommendation. On the other hand, *justification* implies an explanation which is not directly related to the recommendation algorithm, thus it can be generated in a more freely way. Such kind of explanations may be preferred to *transparency* because of algorithms that are difficult to explain or have not to be spread.

The main advantages users may get from the explanation are described in [17] and they include: *transparency*, *scrutability*, *trust*, *effectiveness*, *persuasiveness*, *efficiency* and *satisfaction*. In [19], the authors show how they can be exploited as evaluation metrics for explanatory services. However, providing effective explanations is not always a trivial task; RSs have surely proven to be very accurate in accomplishing their tasks, but they usually work just as black boxes, being not transparent at all. In order to overcome this issue, new methods have been developed in order to generate an explainable recommendation ([18] provides an overview of the most successful approaches proposed over the years) such as *MoviExplain* [16], which exploits movies metadata to justify its recommendation lists. Other interesting works include: a RS based on Restricted Boltzmann Machines which looks at the rating distribution to identify the most explainable items [1], a Latent Factor Model leveraging users reviews to compute more transparent recommendations [23] and, finally, a novel approach based on movies information encoded in the Linked Open Data cloud which generates natural language explanation for the computed recommendation presented in [10]. Looking at the last mentioned method, it is worth noticing how Knowledge Graphs (KG) are recently being used in lots of applications; they freely offer a large amount of structured data which turned out to be very useful also in recommendation scenarios [4, 5, 12]. In particular, in [2], the authors introduce the idea of a Semantics-Aware Autoencoder which paves the way to compute explanation by leveraging deep learning techniques.

As a matter of fact, all the approaches based on deep learning models that have been proposed over the years, turned out to barely leverage on latent factors to which no meaning can be attached to. Among them, Autoencoder Neural Networks have proven their effectiveness in CF settings as shown in [14], in which the authors use an Autoencoder fed with user ratings in order to predict missing value for users' unseen items. In other works such as [22] a stacked architecture made of Autoencoders is proposed to perform a generalization over higher set of latent features that every stacked autoencoder is able to learn. More recently, in [21] the authors propose an hybrid architecture for Autoencoders in order to incorporate both users' feedbacks and content description about items. A similar approach has been proposed in [6], in which they exploit side information in a CF setting by using Stacked Autoencoders in order to overcome the cold start problem and data sparsity.

## 3 BACKGROUND TECHNOLOGIES

Our approach relies on two main technologies: Knowledge Graphs and Autoencoder Neural Networks. The proposed method shows how to use the former to map KG's connections to the topology of the latter, in order to give an explicit meaning to the connections in the NN.

### 3.1 Knowledge Graphs

In the past few years, we have assisted to the publication of freely ontological data on the Web, thanks to diverse communities that began to develop Knowledge Graphs as well-structured graph data encoding the human knowledge. KGs are oriented graphs in which nodes identify resource entities and edges provide labeled relationships between them. Some prominent examples of KGs are DBpedia and Wikidata, which are community driven projects that leverage on Wikipedia pages to automatically parse structured data. Mainly, two kind of information exist in DBpedia: semantics-aware and factual one. The former can be divided into categorical and ontological data. Categorical information is encoded through the dct:subject predicate and represents items categories parsed from Wikipedia infoboxes, such as `Vigilante films` or `Cyborg films`. Categories in Wikipedia are collaboratively maintained by community's editors thus leading to a rich set of categories that reflects a human classification by encoding knowledge about classes attributes and other semantic relations [11]. On the other hand, ontological data basically captures entities types (classes) and their hierarchy; it does not only represent their taxonomy but extends it by using restrictions on its relationships to other classes or on the properties a particular class is allowed to posses. Finally, factual knowledge is merely made of *facts*; it identifies items' attributes, as it can be in the movies domain that the actor `Will Smith` starred in the movie `I, Robot`, as depicted in Figure 1. Differently from categorical information, factual one is identified via different attributes/predicates connecting an item to different entities as for the case of `director`, `starring`, etc..

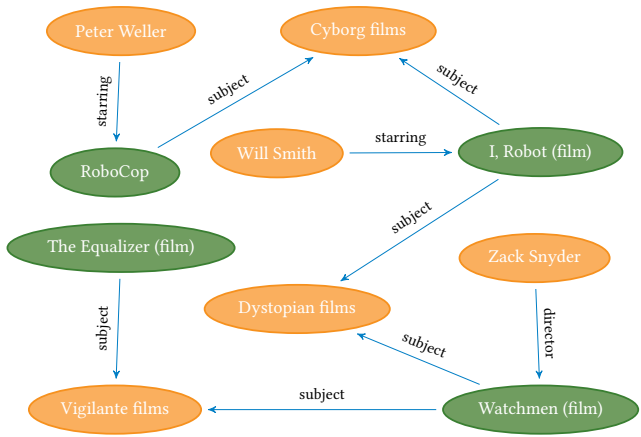

**Figure 1: Part of a KG related to the movie domain.**

## 3.2 Semantics-Aware Autoencoders

Neural Networks model are generally made by one input layer, one or more hidden layers and an output layer. Every layer contains neurons, and every neuron of layer $i$ is connected to all neurons of layer $i + 1$. In particular, Autoencoders are a special kind of unsupervised learning Neural Networks that learn a function able to reconstruct the original data available at the output layer. In the training phase, autoencoders learn how to reconstruct the input vector $x$ through a latent representation encoded in the hidden layers.

In a semantics-aware autoencoder, the hidden layers and their connections are substituted by the knowledge graph thus having an explicit representation on the meaning associated both to hidden nodes and to their mutual connections [2]. This means that each neuron represents an entity in the adopted KG and the edge between two autoencoder nodes exist if the corresponding KG entities are connected with a predicate (labeled edge).

In our implementation, we adopted three different configurations based on a single hidden layer semantics-aware autoencoder (see Figure 2) which exploits one of the following sets of information available in DBpedia: (i) *semantic data*, or rather categorical attributes of items; (ii) *factual data*, specific items properties (such as `actors` and `directors` in the movie domain); (iii) *semantic and factual information*, a mixture of the previous ones.

Hence, the resulting autoencoder has three layers: input layer, hidden layer and output layer where the input and output layers represent items in the catalog while the middle hidden layer contains their DBpedia categories and/or properties.

As previously said, in the training phase, an autoencoder learns how to reconstruct the input vector (in our case user ratings) using the latent representation encoded in the hidden layer. As we train an autoencoder per user, once the model converges, in a semantics-aware autoencoder, for each user we have a latent representation of item's features which, actually, result to be no more latent because every neuron corresponds to an entity in the KG. It turns out that features belonging to positively rated items tend to have a higher weight, differently from those of negatively rated items. This behavior is quite understandable considering that a rating feeding an input node (representing an item in the catalog) flows throughout the neural network by crossing only features/nodes connected to it in the KG. We want to stress here that, although each autoencoder is trained over a not huge number of samples, in [3] we prove that recommendation results have very good performance in terms of accuracy and diversity also compared to state-of-the-art algorithms[3].

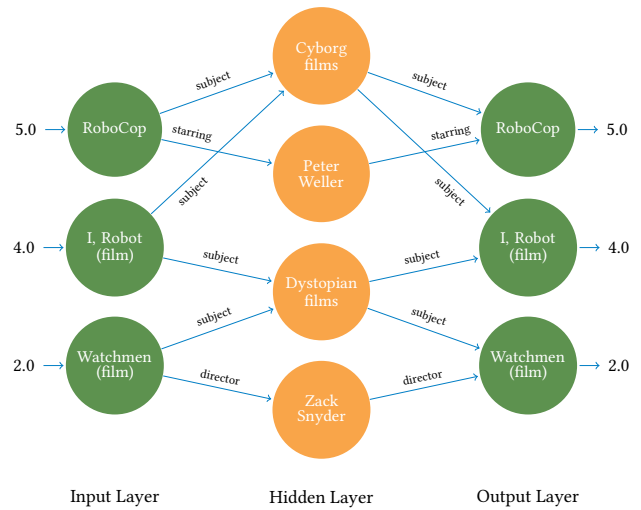

**Figure 2: Architecture of a semantic autoencoder.**

To train such a kind of autoencoder, we inhibit the feedforward and backpropagation step for those neurons which result to be not connected in the KG by using a masking multiplier matrix $M$ where rows and columns represent respectively items and features.

$$M_{m,n} = \begin{pmatrix} a_{1,1} & a_{1,2} & \cdots & a_{1,n} \\ \vdots & \vdots & \ddots & \vdots \\ a_{m,1} & a_{m,2} & \cdots & a_{m,n} \end{pmatrix} \quad (1)$$

---

The matrix in Equation (1) represents the adjacency matrix of the KG where a generic entry is a binary value indicating whether a connection among entities exists in it. In other words, we have

$$a_{i,j} \in M_{m,n} = \begin{cases} 1, & \text{if item } i \text{ is connected to feature } j \\ 0, & \text{otherwise} \end{cases}$$

Hence, hidden ($h$) and output ($o$) layers are computed by the following two equations:

$$h = g(X \times (W_1 \circ M))$$
$$o = g(h \times (W_2 \circ M^T))$$

During the backpropagation step, gradients are computed as usually for $W_2$ and $W_1$ with respect to a mean squared error loss $E = \frac{1}{2}\sum_i \parallel x_i - y_i \parallel^2$ being $x_i$ and $y_i$ the elements of the input and output vector respectively.

The weights update step in SGD (Stochastic Gradient Descent) backpropagation has been modified according to Equations (2) in order to take into account the masking matrix:

$$W_1 = (W_1 \circ M) - r \cdot \frac{\partial E}{\partial W_1} \qquad W_2 = (W_2 \circ M^T) - r \cdot \frac{\partial E}{\partial W_2} \qquad (2)$$

Where $E$ is the mean squared error loss while $W_1$ and $W_2$ represent the weight matrices for the connections between the input and hidden layer ($W_1$) and between the hidden layer and the output layer ($W_2$). They are both initialized randomly using Xavier initialization [7]. In our experiments, we trained the model for 1000 epochs with a learning rate $r = 0.03$ and we used the well-known sigmoid $\sigma(z) = \frac{1}{1+e^{-z}}$ as activation function. Since we train one autoencoder per user and we want it to overfit on user ratings, we did not use any form of regularization.

**Computing user profiles.** After training the autoencoder for each user $u$, we extract the weights of the hidden neurons and use them to build a user profile $P(u)$:

$$P(u) = \{\langle f_{u1}, w_{u1}\rangle, \ldots, \langle f_{um}, w_{um}\rangle\}$$

being $f_u$ the label associated to the neuron and $w_u$ its corresponding weight for $u$. Indeed, as each hidden neuron represents an entity in DBpedia, we may assume that its weight after the training is an indicator of the importance of the corresponding entity for $u$.

## 4 COMPUTING SEMANTICS-AWARE EXPLANATIONS

As previously said, in this paper we explore the adoption of a semantic autoencoder to provide an explanation for top-$N$ recommendations. In our experimental setting aimed at evaluating the explainability of the trained model, while building the structure of the SemAuto autoencoder we used those KG entities reachable through the predicate `dct:subject` as item categories, while we used the approach originally proposed in [13] to select the top-3 factual movie properties: `dct:starring`, `dct:director`, `dct:writer`[4].

[4]We selected only the top-3 properties to reduce the dimension of the feature space and then minimize the noise in the provided explanation. Finding the best number of properties to compute explanations is not in the scope of this paper and is part of our future work.

In order to formulate a human-understandable explanation for the provided results, we rely on the weights associated to features in the user profile, which also appear in the description of the recommended items. In particular, given a user $u$ and a recommendation list $rec(u) = [\langle i_1, \tilde{r}_1^u\rangle, \ldots, \langle i_n, \tilde{r}_n^u\rangle]$, with $\tilde{r}_k^u$ being a score/rating computed for the item $i_k$ by a recommendation engine, we may compute a pointwise and a pairwise personalized explanation.

**pointwise personalized.** Given an item $i = \{f_{1i}, f_{2i}, \ldots, f_{ni}\}$ described by a set of features $f_i$, the pointwise explanation $e1^k(i)$ is computed by considering the set of top-$k$ highest weighted features in $P(u)$ which also appear in $i$.

**pairwise personalized.** Given two items $i$ and $j$ such that $\tilde{r}_i^u > \tilde{r}_j^u$, the pairwise explanation $e2^k(i, j)$ is computed by evaluating both $e1^k(i)$ and $e1^k(j)$. In case $m$ features are in common between $e1^k(i)$ and $e1^k(j)$, we compute $e1^{k+m}(j)$ and leave them only in $e1^k(i)$ thus avoiding any overlap between the explanation for $i$ and that for $j$.

To verify that the explanation generated through a Semantics-Aware Autoencoder is able to satisfy the main explanatory criteria of *transparency*, *persuasiveness*, *effectiveness*, *trust* and *satisfaction*, we built a web platform[5] that returns the top-5 recommendations and then asks for users' feedback about the provided explanation.

### 4.1 Explanation styles

We provided our platform with four different explanation styles: as in [10], we used a *popularity-based explanation* and a *non-personalized* one as baselines [19]. As a third style we propose our pairwise approach. During the usage of the platform by a user, we randomly select one of the three styles and show the associated explanation, which is generated for the top-2 recommended items in a *pairwise* fashion. Hence, the user may receive one of the following explanations:

**popularity-based** *We suggest these items since they are very popular among people who like the same movies as you.*

**(non-/pointwise) personalized** *We guess you would like to watch $i$ and $j$ since they are about $\hat{f}_{u1}, \ldots, \hat{f}_{uk}$* (Example 4.1)

**pairwise personalized** *We guess you would like to watch $i$ more than $j$ because you may prefer $e1^k(i)$ over $e1^{k+m}$($j$)* (Example 4.1)

*Example 4.1.* In order to show the difference between a pointwise and a pairwise personalized explanation, hereafter we report the two explanation styles with reference to a recommendation having *Terminator 2: Judgment Day* and *Transformers: Revenge of the fallen* as the first two items in the recommendation list. The pointwise personalized explanation may look like:

*We guess you would like to watch Terminator 2: Judgment Day (1991) and Transformers: Revenge of the Fallen (2009) because you may prefer:*

- *(subject) 1990s science fiction films*
- *(subject) Science fiction adventure films*
- *(subject) Drone films*

[5]Available at **http://sisinflab.poliba.it/semanticweb/lod/recsys/explanation**

- *(subject) Cyberpunk films*

*and:*

- *(subject) Science fiction adventure films*
- *(subject) Films set in Egypt*
- *(subject) Robot films*
- *(subject) Films shot in Arizona*
- *(subject) Ancient astronauts in fiction*

while the pairwise version (see also Figure 3b) is a bit different:

> *We guess you would like to watch Terminator 2: Judgment Day (1991) more than Transformers: Revenge of the Fallen (2009) because you may prefer:*

- *(subject) 1990s science fiction films*
- *(subject) Science fiction adventure films*
- *(subject) Drone films*
- *(subject) Cyberpunk films*

*over:*

- *(subject) Films set in Egypt*
- *(subject) Robot films*
- *(subject) Films shot in Arizona*
- *(subject) Ancient astronauts in fiction*
- *(subject) IMAX films*

The *popularity-based explanation* may be considered as the less meaningful, since it justifies recommender choices by just leveraging the popularity of suggested items among the users with similar tastes of the active user $u$. The *non-personalized explanation*, instead, tries to explain the provided recommendation by using additional information about the suggested items. In our experiments, we randomly select $k = 5$ features from the set $F_{ij} = F_i \cup F_j = \{f_{1i}, f_{2i}, \ldots, f_{ni}\} \cup \{f_{1j}, f_{2j}, \ldots, f_{n'j}\}$. In a similar manner, in a *pointwise personalized explanation* we selected the top-5 features from each set $F_i$ and $F_j$. The value $k = 5$ has been selected also to compute $e2^k(i, j)$ in the *pairwise personalized explanation*.

Please notice that the considered set of features per item varies according to the different configuration adopted for the `SemAuto` autoencoder; it may include just item categories, factual data or both of them.

## 4.2 Evaluation Protocol

During the online A/B testing phase, we fixed a sequence of steps in order to measure the aforementioned explanatory criteria.

**Steps 1-3.** At the beginning of the experiment, the user $u$ selects at least 15 movies she has watched among the ones randomly listed by the platform. The movies belong to the well-known MovieLens 20M dataset [6]. Then, she is invited to rate each selected movie on a five-stars rating scale; data so gathered are exploited to get both the user profile computed with the semantic autoencoder and a top-5 recommendation list.
**Step 4.** Once the recommendation has been generated, the user is asked to rate the suggested items, even if no explanation has been

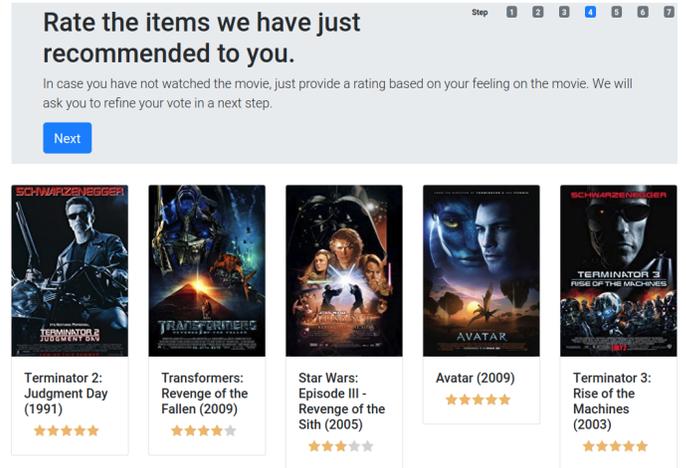

(a) Step 4. The user is asked to rate the recommended items, even if she has not watched them.

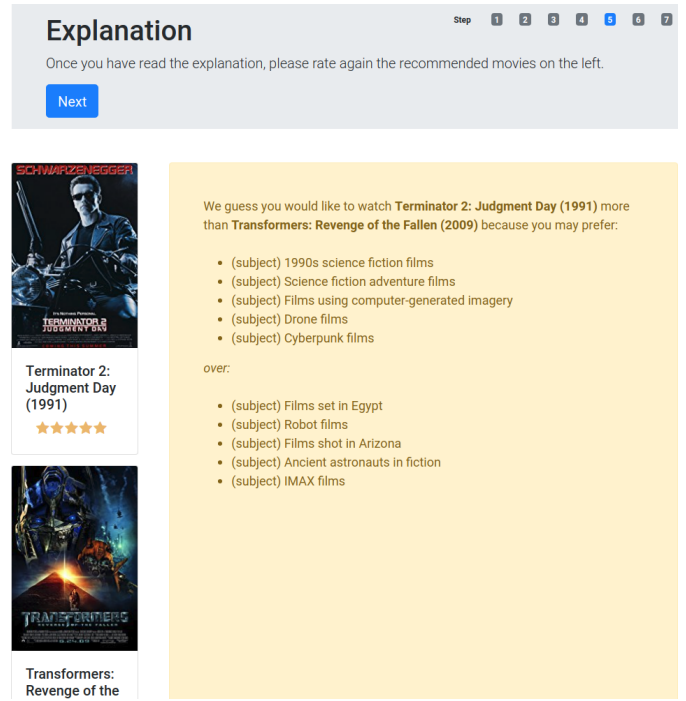

(b) Step 5. The user is asked to read the explanation and after that to rate again the top-2 recommended items.

Figure 3: Screenshots of the A/B testing platform.

shown yet: these ratings will be relevant to determine the impact the explanation has on the user (*persuasiveness*).
**Step 5.** The next step consists of showing to $u$ one of the three randomly selected explanation styles deployed within the application. After enjoying the explanation, the user has to re-rate the top-2 recommended items, letting us measure how different is the items evaluation before and after the explanation has been provided.

**Step 6.** Similarly, in the last part of the experiment, the user is asked to re-rate the recommended movies after watching the related trailers. This phase allows $u$ to emulate the items consumption, and makes her more aware about the topics of the suggested movies. In this way we can evaluate how much effective the selected explanation style was (*effectiveness*).

**Step 7.** Finally, the user fills a questionnaire, aimed at measuring the explanation *transparency*, *trust* and *satisfaction* (see Table 1).

## 4.3 Metrics

When evaluating an explanation system, the main characteristics to evaluate are [18]:

- *transparency*, which refers to the capability of the explanation to make users aware of how the system works;
- *trust*, or rather the confidence users have in the system;
- *satisfaction*, if users have an enjoyable experience in the usage of the system;
- *persuasiveness*, which evaluates how much convincing is the proposed explanation;
- *effectiveness*: the explanation is said to be effective if it helps users to correctly estimate items relevance before the consumption.

The first three characteristics are evaluated by collecting answers from users after filling the questionnaire at Step 7. As a final score for the first and the second metric we used the percentage of users that answered positively to the questions, while we exploited the average score assigned by users to quantify the overall *satisfaction*.

In order to evaluate the *persuasiveness* of the proposed explanation, we asked users to rate each recommended item before and after showing them the explanation: if the rating provided after looking at the explanation is higher than the original one, then the explanation has been able to persuade the user to try the suggested item. More formally we measure persuasiveness as [18]:

$$persuasiveness = \frac{1}{|U|} \cdot \sum_{u \in U} \frac{1}{N} \cdot \sum_{i \in I_N^u} (r_{ui}^e - r_{ui})$$

where $U$ stands for the collection of users; $I_N^u$ represents the set of top-$N$ recommended items for $u$; $r_{ui}$ and $r_{ui}^e$ are, respectively, the ratings $u$ assigns to $i$ just before and after the explanation is provided.

Analogously, we evaluated the *effectiveness* as the difference between two ratings (see Equation (4.3) [18]), where $r_{ui}^t$ represents the rating the user gives to the suggested movie after watching the related trailer ($r_{ui}^t$).

$$effectiveness = \frac{1}{|U|} \cdot \sum_{u \in U} \frac{1}{N} \cdot \sum_{i \in I_N^u} ||r_{ui}^e - r_{ui}^t||$$

The lower this value, the more effective the explanation, since it implies that users have rated each item with very similar values before and after the explanation has been provided.

| METRIC | QUESTION |
|---|---|
| *transparency* | I understood the reason why the two movies have been ranked in the proposed order. |
| *trust* | The explanation increased my trust in the system. |
| *satisfaction* | The provided explanation: **really** captures my tastes. **partially** captures my tastes. **does not** capture my tastes. |

**Table 1: The final questionnaire.**

## 5 RESULTS DISCUSSION

We conducted our experiment with the help of 892 volunteers, with at least 73 subjects for each of the implemented settings. As stated in [9], 73 has to be considered as the minimum acceptable sample size for such kind of experiments. This assures the significance of our experimental results. Furthermore, we verified the statistical significance of our experiment by using *Wilcoxon Rank-Sum Test*, getting $p \ll 0.01$.

As shown in Figure 4, a content-based explanation is always preferred by users, since the *popularity-based* style gets the worst results in all the considered explanatory criteria. The only exception is represented by *persuasiveness*: quite interestingly, the *non-personalized* explanation leveraging semantic/categorical information gets negative values, as it happens when users rate items with lower values after looking at the explanation than before, being not convinced to consume the suggested items at all. Hence, users overestimate their interest in the recommended items or underestimate it because of the provided explanation; this may be interpreted as users dissatisfaction for the shown categories, since they are chosen randomly without taking into account users interests. As a matter of fact, the *personalized* approach, which computes the explanation by leveraging users preferences, outperforms all its competitors, even with the pairwise approach. Furthermore, it is worth noticing how convincing the categories are: by looking at the results, if the *personalized* style exploits categorical features, then it performs very well in terms of *persuasiveness* if compared to others. It is worth noticing that when categorical features are combined with factual information, they lead to a better *persuasiveness*, in particular the pairwise approach gets better results than the pointwise one. This may be explained by considering that we simulate item consumptions through their associated trailers: in our experiments, users provide a certain rating to a movie by just considering a few scenes, that are those shown in the trailer. Therefore, users may get information about the movie topics, subjects and how good or interesting an actor's or a director's performances are. This condition may influence the way the explanation is perceived by users, who demonstrated to find more convincing an explanation involving both categorical and factual information rather than an explanation based on item factual properties only.

On the other hand, still considering the pairwise *personalized* approach, factual properties turn out to be more effective as concerning *satisfaction*, *trust* and *effectiveness*; we suppose that users feel more confident in specific information such as actors or directors rather than just a set of movie categories. This trend is already confirmed by the pointwise approach, and the pairwise one gets even higher score with those metrics. As a matter of fact, the system *transparency* has the highest values when both semantic and factual properties are exploited; as these values are very close to those achieved by just using factual properties, we can claim that it is the factual information itself that improves the measured performances. Analogously, richer items descriptions make more *effective* the explanation: when the system leverages both categories and factual attributes, the *effectiveness* achieves its best results. Hence, providing more information about the suggested items surely lets users better evaluate them before their consumption. By considering the *non-personalized* style, it is quite interesting that by using both categorical and factual attributes, the gathered results for all the adopted criteria are usually the best, far from the performances measured by the other settings based on the KG. Once again, as discussed above, this may depend on the random aspect behind it: e.g., users may be more or less *satisfied* with the provided explanation according to the randomly shown features, which they may like or dislike, know or ignore. Summing up, from the experimental results, we may argue that the pairwise approach with factual information gets better performance in users' *satisfaction*, *effectiveness* and *trust*, while it outperforms the pointwise one in *persuasiveness* and *transparency* when both factual and categorical information are exploited.

To provide an answer to **RQ1**, examining the results, it turns out that our SemAuto provides reliable users' descriptions as evidenced in the *effectiveness* metric which gets the lowest value by using a pairwise explanation. This can be interpreted as a strong signal that the information encoded in the autoencoder hidden layer is representative of the users' preferences because the users is less prone to change her ratings after she read the explanation.

As for **RQ2**, we can assert that the pairwise approach outperforms the pointwise one in all metrics especially in *transparency* because it provides a better justification on how the system ranks items according to the importance of the features in the user profile. This lets the user to better understand how her preferences are involved in the recommendation process. In fact, this has an impact especially for the *persuasiveness* metric where the pairwise approach has a higher score with respect to the pointwise explanation, thus leading users in consuming an item after they have read the provided explanation.

## 6 CONCLUSION AND FUTURE WORK

In this paper we present results on the capability for a semantics-aware autoencoder [2] to generate explanation to recommendation lists via the exploitation of data coming from the DBpedia knowledge graph. Online experimental results show that a content based explanation is preferred by users, as it outperforms other baselines in terms of transparency, trust, satisfaction, persuasiveness and effectiveness. As we can see in the *satisfaction*, *effectiveness* and *trust* plots for both pointwise and pairwise approaches, an interesting point is that, in order to build an explanation, factual data works better than the semantic/categorical one, achieving the same results as when both semantic and factual data are exploited. A possible reason for this behavior is that the probability for a user to know factual data and accepts it as explanation is higher if compared to categorical one. Very interestingly, a pairwise approach has the same trends for all the evaluation metrics of the pointwise one but it outperforms the latter.

As future work it would be interesting to investigate about the system's *scrutability*, by allowing users to correct the recommender engine reasoning. Explanations here should be part of a continuous cycle where the user understands how the system is working under the hood and takes control over the type of recommendations made by the engine. This continuous loop could pave the way to a new kind of conversational recommender systems in which the user is allowed to explore and move in the feature space by knowing which features relevant to her are involved in the recommendation process.

## REFERENCES

[1] Behnoush Abdollahi and Olfa Nasraoui. 2016. Explainable Restricted Boltzmann Machines for Collaborative Filtering. *CoRR* abs/1606.07129 (2016).
[2] Vito Bellini, Vito Walter Anelli, Tommaso Di Noia, and Eugenio Di Sciascio. 2017. Auto-Encoding User Ratings via Knowledge Graphs in Recommendation Scenarios. In *Proceedings of the 2nd Workshop on Deep Learning for Recommender Systems*. ACM, 60–66.
[3] V. Bellini, A. Schiavone, T. Di Noia, A. Ragone, and E. Di Sciascio. 2018. Computing recommendations via a Knowledge Graph-aware Autoencoder. *ArXiv e-prints* (July 2018). arXiv:cs.IR/1807.05006
[4] Marco de Gemmis, Pasquale Lops, Cataldo Musto, Fedelucio Narducci, and Giovanni Semeraro. 2015. *Semantics-Aware Content-Based Recommender Systems*. Springer US, Boston, MA, 119–159.
[5] Tommaso Di Noia, Roberto Mirizzi, Vito Claudio Ostuni, Davide Romito, and Markus Zanker. 2012. Linked open data to support content-based recommender systems. In *Proceedings of the 8th International Conference on Semantic Systems*. ACM, 1–8.
[6] Xin Dong, Lei Yu, Zhonghuo Wu, Yuxia Sun, Lingfeng Yuan, and Fangxi Zhang. 2017. A Hybrid Collaborative Filtering Model with Deep Structure for Recommender Systems.. In *AAAI*. 1309–1315.
[7] Xavier Glorot and Yoshua Bengio. 2010. Understanding the difficulty of training deep feedforward neural networks. In *Proceedings of the thirteenth international conference on artificial intelligence and statistics*. 249–256.
[8] Jonathan L. Herlocker, Joseph A. Konstan, and John Riedl. 2000. Explaining Collaborative Filtering Recommendations. In *Proceedings of the 2000 ACM Conference on Computer Supported Cooperative Work (CSCW '00)*. ACM, New York, NY, USA, 241–250.
[9] Bart P. Knijnenburg and Martijn C. Willemsen. 2015. *Evaluating Recommender Systems with User Experiments*. Springer US, Boston, MA, 309–352.
[10] Cataldo Musto, Fedelucio Narducci, Pasquale Lops, Marco De Gemmis, and Giovanni Semeraro. 2016. ExpLOD: A Framework for Explaining Recommendations Based on the Linked Open Data Cloud. In *Proceedings of the 10th ACM Conference on Recommender Systems (RecSys '16)*. ACM, New York, NY, USA, 151–154.
[11] Vivi Nastase and Michael Strube. 2008. Decoding Wikipedia Categories for Knowledge Acquisition.. In *AAAI*, Vol. 8. 1219–1224.
[12] Sergio Oramas, Vito Claudio Ostuni, Tommaso Di Noia, Xavier Serra, and Eugenio Di Sciascio. 2016. Sound and Music Recommendation with Knowledge Graphs. *ACM Trans. Intell. Syst. Technol.* 8, 2, Article 21 (Oct. 2016), 21 pages.
[13] Azzurra Ragone, Paolo Tomeo, Corrado Magarelli, Tommaso Di Noia, Matteo Palmonari, Andrea Maurino, and Eugenio Di Sciascio. 2017. Schema-summarization in Linked-data-based Feature Selection for Recommender Systems. In *Proceedings of the Symposium on Applied Computing (SAC '17)*. ACM, New York, NY, USA, 330–335.
[14] Suvash Sedhain, Aditya Krishna Menon, Scott Sanner, and Lexing Xie. 2015. AutoRec: Autoencoders Meet Collaborative Filtering. In *Proceedings of the 24th International Conference on World Wide Web (WWW '15 Companion)*. ACM, New York, NY, USA, 111–112.
[15] Rashmi Sinha and Kirsten Swearingen. 2002. The Role of Transparency in Recommender Systems. In *CHI '02 Extended Abstracts on Human Factors in Computing Systems (CHI EA '02)*. ACM, New York, NY, USA, 830–831.
[16] Panagiotis Symeonidis, Alexandros Nanopoulos, and Yannis Manolopoulos. 2009. MoviExplain: A Recommender System with Explanations. In *Proceedings of the Third ACM Conference on Recommender Systems (RecSys '09)*. ACM, New York, NY, USA, 317–320.
[17] Nava Tintarev and Judith Masthoff. 2007. A Survey of Explanations in Recommender Systems. In *Proceedings of the 2007 IEEE 23rd International Conference on Data Engineering Workshop (ICDEW '07)*. IEEE Computer Society, Washington, DC, USA, 801–810.

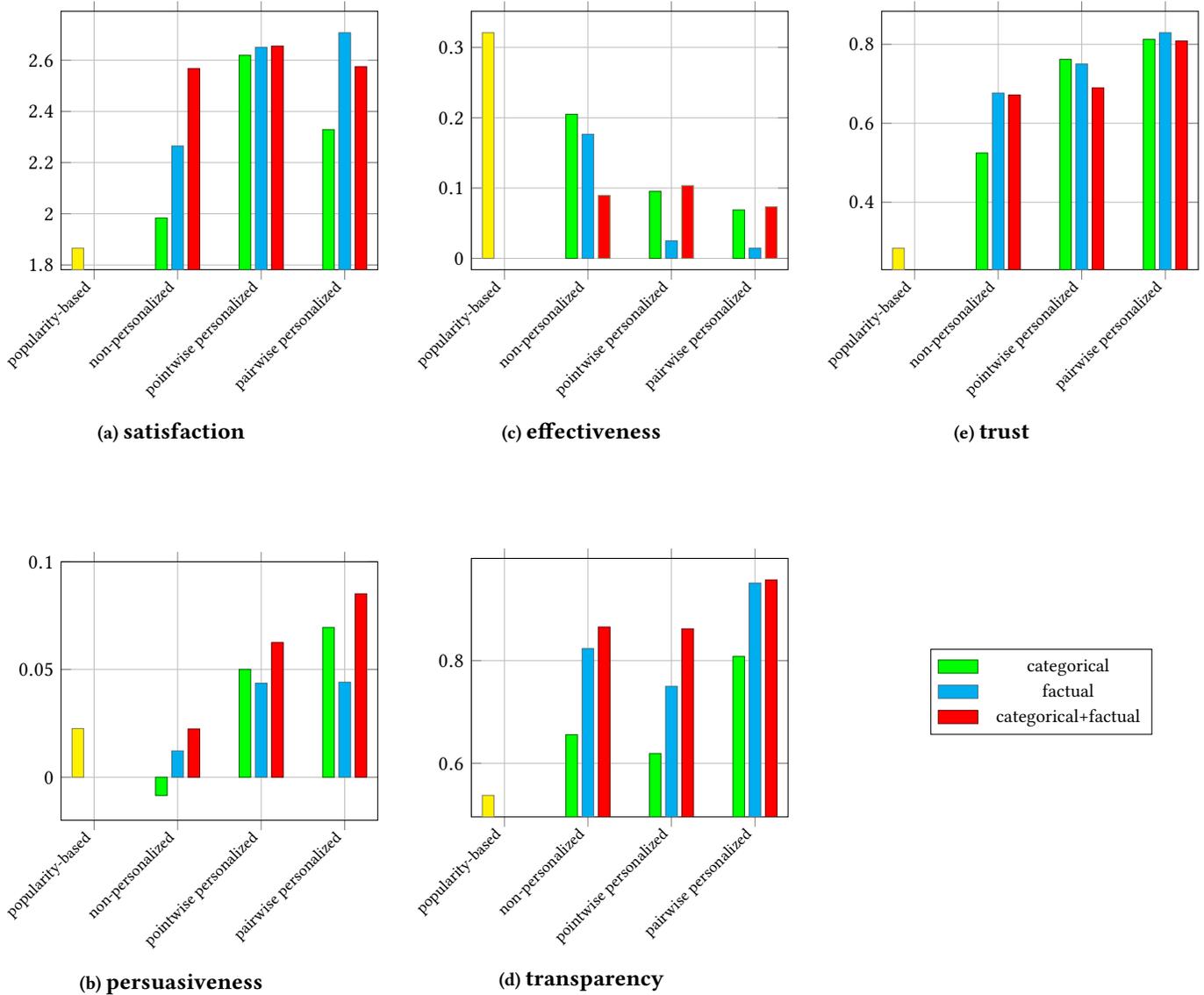

(a) satisfaction

(c) effectiveness

(e) trust

(b) persuasiveness

(d) transparency

■ categorical
■ factual
■ categorical+factual

Figure 4: Results comparison.

[18] Nava Tintarev and Judith Masthoff. 2011. *Designing and Evaluating Explanations for Recommender Systems.* Springer US, Boston, MA, 479–510.

[19] Nava Tintarev and Judith Masthoff. 2012. Evaluating the Effectiveness of Explanations for Recommender Systems. *User Modeling and User-Adapted Interaction* 22, 4-5 (Oct. 2012), 399–439.

[20] Jesse Vig, Shilad Sen, and John Riedl. 2009. Tagsplanations: Explaining Recommendations Using Tags. In *Proceedings of the 14th International Conference on Intelligent User Interfaces (IUI '09).* ACM, New York, NY, USA, 47–56.

[21] Hao Wang, Xingjian SHI, and Dit-Yan Yeung. 2016. Collaborative Recurrent Autoencoder: Recommend while Learning to Fill in the Blanks. In *Advances in Neural Information Processing Systems 29*, D. D. Lee, M. Sugiyama, U. V. Luxburg, I. Guyon, and R. Garnett (Eds.). Curran Associates, Inc., 415–423.

[22] Hao Wang, Naiyan Wang, and Dit-Yan Yeung. 2015. Collaborative Deep Learning for Recommender Systems. In *Proceedings of the 21th ACM SIGKDD International Conference on Knowledge Discovery and Data Mining (KDD '15).* ACM, New York, NY, USA, 1235–1244.

[23] Yongfeng Zhang, Guokun Lai, Min Zhang, Yi Zhang, Yiqun Liu, and Shaoping Ma. 2014. Explicit Factor Models for Explainable Recommendation Based on Phrase-level Sentiment Analysis. In *Proceedings of the 37th International ACM SIGIR Conference on Research & Development in Information Retrieval (SIGIR '14).* ACM, New York, NY, USA, 83–92.